\scrollmode

\documentclass[a4paper,12pt,prb]{revtex4}
\usepackage{graphics}
\usepackage{amssymb}
\usepackage[a4paper]{geometry}
\usepackage{url}
\usepackage{setspace}

\begin{document}

\title{Self-Consistent Ornstein-Zernike approximation for the Yukawa fluid
with improved direct correlation function}

\author{A.~Reiner$^1$ and J.~S.~H{\o}ye$^2$,\\
Institutt for fysikk,\\
Norges teknisk-naturvitenskapelige universitet (NTNU),\\
N-7491 Trondheim, Norway.\\
{}$^1$e-mail: {\tt areiner@tph.tuwien.ac.at}\quad
{}$^2$e-mail: {\tt johan.hoye@ntnu.no}}

\begin{abstract}%%
Thermodynamic consistency of the Mean Spherical Approximation as well
as the Self-Consistent Ornstein-Zernike Approximation (SCOZA) with the
virial route to thermodynamics is analyzed in terms of renormalized
$\gamma$-ordering.  For continuum fluids this suggests the addition of
a short-range contribution to the usual SCOZA direct correlation
function, and the shift of the adjustable parameter from the potential
term to this new term.  The range of this contribution is fixed by
imposing consistency with the virial route at the critical point.
Comparison of the results of our theory for the hard-core Yukawa
potential with simulation data show very good agreement for cases
where the liquid-vapor transition is stable or not too far into the
metastable region with respect to the solid state.  In the latter case
for extremely short-ranged interactions discrepancies arise.
\end{abstract}

\maketitle

\newpage

\section{Introduction}

By use of the Self-Consistent Ornstein-Zernike Approximation (SCOZA),
very accurate results have been obtained for the equation of state for
fluids \cite{scoza:4,scoza:6} and lattice systems (or the Ising spin
system) \cite{scoza:20,scoza:8,scoza:21,scoza:26}.  Also in the
critical region the results are very accurate, showing non-classical
critical behavior \cite{scoza:20}, but a form of generalized scaling
is obtained in the critical region instead of full scaling
\cite{scoza:25}.

Looking more closely into the data one finds that for the lattice gas,
or nearest neighbor Ising model, results are somewhat better than
those for the continuum fluid.  The SCOZA critical temperature $ T_c$
of the former is about $0.2\,\%$ away from the best estimates
\cite{scoza:1}, while this separation is about $0.6\,\%$ for the
commonly considered Yukawa fluid ($z=1.8$) according to recent
estimates based on Monte Carlo (MC) computations
\cite{scoza:6,hc-y:6}.  Here we will analyze this situation in view of
renormalized $\gamma$-ordering \cite{hoye:dr,hoye::nato}, and we will
propose a modified version of SCOZA.  A slightly simplified variant of
this scheme is implemented and solved numerically, yielding
essentially perfect agreement with simulations for potential ranges
relevant for the liquid state.  For extremely short-ranged potentials
where the gas-liquid transition is deeply buried inside the metastable
region, however, agreement is less satisfactory, and the numerical
procedure may even fail to converge to consistency with the virial
pressure.

As SCOZA builds upon the Mean Spherical Approximation (MSA), we can
understand that SCOZA is more accurate for lattice gases than for
fluids by noting that this is true for the MSA, too.  By further
analysis it is seen that for fluids a significant contribution to the
direct correlation function of short range is missing.  When this
contribution is included, it enlarges the amplitude of the correlation
function and thus amplifies the internal energy due to correlations
(which is negative).  This further increases the deviation from the
corresponding mean field or Van der Waals type equation of state.
Since correlations lower the critical temperature, the net result is a
further lowering of it.  For the Yukawa interaction typically used as
a model potential in simulations, we expect this additional lowering
to be something like $(0.5\pm0.4)\,\%$ as a crude estimate.  An
immediate problem here is the determination of the proper range of
this short range piece of the direct correlation function.  In this
respect we can be guided by considering thermodynamic consistency with
the virial theorem to estimate the range mentioned.

In section~\ref{sec:gamma} we will investigate the relation between
the MSA and renormalized $\gamma$-ordering for potentials of long
range.  Then in section~\ref{sec:scoza} thermodynamic consistency with
the virial theorem is considered and applied to SCOZA.  A modified
theory is then formulated on this basis in section~\ref{sec:theory},
and applied to the one-Yukawa problem numerically in
section~\ref{sec:num}.  A short summary (section~\ref{sec:summary})
concludes our contribution.

\section{Renormalized $\gamma$-ordering and MSA}

\label{sec:gamma}

The MSA is the solution of the Ornstein-Zernike (OZ) equation,
\begin{displaymath}
\tilde h(k) = \frac{\tilde c(k)}{1-\rho\,\tilde c(k)},
\end{displaymath}
with boundary conditions
\begin{equation} \label{eq:closure:msa}
\begin{array}{rll}
\displaystyle h(r)&\displaystyle = -1&\displaystyle \hbox{for }r<1,\\
\displaystyle c(r)&\displaystyle = -\beta\psi(r)&\displaystyle \hbox{for
}r>1.
\end{array}
\end{equation}
Here $h(r)$ is the pair correlation function, $c(r)$ is the direct
correlation function, and
\begin{displaymath}
\psi(r) = -e^{-z\,(r-1)}/r\qquad (r>1),
\end{displaymath}
is the perturbing interaction; the hard core diameter and the unit of
energy are set to unity for simplicity; and a tilde denotes Fourier
transformation.  (It should be noted that $\psi(r)$ is not restricted
to interactions of Yukawa form.)

For a lattice gas or an Ising spin system the boundary conditions
become
\begin{displaymath}
\begin{array}{rll}
\displaystyle h(0)&\displaystyle = -1,\\
\displaystyle c(\vec r)&\displaystyle = -\beta\psi(\vec r)&\displaystyle
\hbox{for }\vec r\ne0.
\end{array}
\end{displaymath}
So in this case we have
\begin{equation} \label{eq:lattice:core}
\rho + \rho^2h(0) = \rho (1-\rho)
= \frac1{(2\pi)^3}\int\frac\rho{1-\rho\tilde c(k)}{\mathrm{d}^3} k
\end{equation}
for lattice cells of unit volume.  (To keep the notation simple, we do
not explicitly indicate that $k$ and similar arguments are necessarily
vectors in the lattice case, nor do we explicitly point out
restriction to the Brillouin zone, \textit{i.~e.{}}, $-\pi \le k_i \le
\pi$ with $i = x,y,z$.)  Defining $\tilde v(k) = -\beta\tilde\psi(k)$,
we can now write
\begin{displaymath}
\tilde c(k) = c_0 + \tilde v(k)
\end{displaymath}
where $c_0$ is determined by Eq.~(\ref{eq:lattice:core}).  For the
reference system alone we then have ($\tilde\psi(k)=0$ by which $h(0)=\tilde h(k)$)
\begin{equation} \label{eq:def:mu}
\mu =\rho+\rho^2 \tilde{h}(k)\equiv \rho(1-\rho) = \frac\rho{1-\rho c_0}.
\end{equation}
More generally for $\tilde\psi(k) \ne 0$ we can define
\begin{displaymath}
\mu_c = \frac\rho{1-\rho c_0},
\end{displaymath}
and the core condition~(\ref{eq:lattice:core}) becomes
\begin{equation} \label{eq:core:mu}
\mu = \mu_c + \mu_c^2K
\end{equation}
with
\begin{displaymath}
K = \frac1{(2\pi)^3}\int\frac{\tilde v(k)}{1-\mu_c\tilde
v(k)}{\mathrm{d}^3} k.
\end{displaymath}

With $\mu$ given by Eq.~(\ref{eq:def:mu}), relation~(\ref{eq:core:mu})
can also be written as
\begin{equation} \label{eq:core:renorm}
\mu = \mu_c (1-\frac12\mu''\mu_c K),
\end{equation}
where a prime denotes differentiation with respect to $\rho$.  In this
form we can recognize the relation obtained by the resummation or
renormalization process to obtain the equation of state for fluids
when the inverse range of the attractive interaction $\gamma$ is used
as a perturbing parameter \cite{hoye:dr,hoye::nato} (see Eqs.~(26)
and~(36) of Ref.~\onlinecite{hoye::nato}).  In Eq.~(\ref{eq:core:renorm}),
$\mu_c$ represents the renormalized hypervertex $\mu$ when graph
expansion in terms of $\gamma$-ordering is considered
\cite{hemmer:1964,lsb:1965}.  The $\mu$ follows from the reference
system, \textit{e.~g.{}}, hard spheres, by use of the compressibility
relation as
\begin{equation} \label{eq:jh:10}
\mu = \rho \left(\frac{\partial\beta p_0}{\partial\rho}\right)^{-1}
= \rho + \rho^2\tilde h_0(0),
\end{equation}
where $p_0$ and $h_0(r)$ are the pressure and the correlation function
of the reference system.  Thus $\mu$ represents the Mayer graphs of
the correlation function of the reference system.  For small $\gamma$
they are regarded as $\delta$-functions of relative distance $r$ so
that only the integral~(\ref{eq:jh:10}) is needed as a leading order
approximation.

Now with renormalization~(\ref{eq:core:renorm}) the equation of state
(EOS) becomes more consistent, with better agreement between the
energy and compressibility routes and essentially common critical
point \textit{via}\ the two routes~\cite{hoye:dr}.  But the remaining
thermodynamic inconsistency prevents well-defined isotherms near the
critical point, which is also the situation of the MSA for the lattice
gas or Ising model.

In contrast to the MSA for lattice systems, the MSA for continuum
fluids is not consistent with Eq.~(\ref{eq:core:renorm}) and therefore
farther away from thermodynamic self-consistency.  In fact, for
interactions of very long range, \textit{i.~e.{}}, small $\gamma$, the
MSA for continuum fluids also fulfills Eq.~(\ref{eq:core:mu}) but not
Eq.~(\ref{eq:core:renorm}).  This is due to the difference in
reference systems.  For a Percus-Yevick hard sphere reference system
one has
\begin{displaymath}
\mu=\frac{(1-\eta)^4}{(1+2\eta)^2}
\end{displaymath}
instead of expression~(\ref{eq:def:mu}), where $\eta=\frac\pi6\rho$
for spheres of unit diameter.  From this $\mu'' =
\partial^2\mu/\partial\rho^2$ can be evaluated at the critical
density, \textit{i.~e.{}}, $\rho_c\approx0.3$ or $\eta_c\approx0.15$
for interactions of realistic range.  (In mean field, $\eta_c=0.129$.)
With $\eta=0.15$ we find
\begin{equation} \label{eq:jh:12}
\frac12\mu'' = -0.53\ldots \approx - 0.5
\end{equation}
which is only one half of the ``MSA value'' that follows from
Eq.~(\ref{eq:core:mu}).

Compared with mean field the finite range of interaction lowers the
critical temperature $ T_c$, and we have
\begin{equation} \label{eq:jh:13}
T_c \approx \frac{\mu_c}\mu T_c^\mathrm{MF},
\end{equation}
where $ T_c^\mathrm{MF}$ is the mean field value.  (The spinodal curve
and thus $ T_c$ follow from the denominator of the integrand of
integral~(\ref{eq:core:mu}) being zero at $k=0$.)  In view of this it
is clear that the MSA compressibility critical point is lowered too
much so that the more accurate energy route gives a classical critical
point well above the former.  For interactions of shorter range where
details of the core condition and hard core correlations become more
important the above results will be modified, but the qualitative
features of the MSA inconsistency will remain.

\section{SCOZA evaluations}

\label{sec:scoza}

The SCOZA differs from the MSA in its imposition of thermodynamic
consistency between the energy and compressibility routes to
thermodynamics: A state-dependent free parameter is introduced into
the pair structure and determined in such a way that the consistency
condition
\begin{equation} \label{eq:scoza:pde}
\frac{\partial a}{\partial \beta}
= \rho\frac{\partial^2\rho u}{\partial\rho^2},
\end{equation}
holds.  Here 
\begin{equation} \label{eq:route:eny}
u = \frac12 \rho \int\psi(r)(h(r)+1){\mathrm{d}^3} r
\end{equation}
is the configurational part of the internal energy per particle as
computed from the energy route, and
\begin{equation} \label{eq:route:comp}
a = 1-\rho \tilde c(0),
\end{equation}
is the reduced inverse compressibility as obtained from the
compressibility route.

In earlier SCOZA evaluations the MSA form~(\ref{eq:closure:msa}) of
the direct correlation function has been used with $\beta$ replaced by
an effective value $\beta_e$.  The contribution to the internal energy
per particle $u_1$ from correlations is then given by
integral~(\ref{eq:core:mu}) as
\begin{equation} \label{eq:scoza:u1}
-\rho u_1 = \frac1{2\beta_e} \mu_c K
\end{equation}
with $\tilde v(k) = -\beta_e\tilde\psi(k)$.  (Note that a
``self-energy'' term $-\rho\psi(0)$ is included in~(\ref{eq:scoza:u1}),
but it will not contribute in~(\ref{eq:scoza:pde}) anyway.)  In
addition one has the mean field term $u_0=\rho^2 \tilde{\psi}(0)/2$ such
that $u=u_0+u_1$.  For continuum fluids $\mu_c$ should also depend
upon $k$, \textit{i.~e.{}}, $\mu_c=\tilde\mu_c(k)$, which we disregard
here to simplify the argument.  Now at the critical point
$1-\mu_c\beta_{ec}(-\tilde\psi(0)) = 0$ is valid both for MSA and
SCOZA that builds directly upon MSA \textit{via}\
Eq.~(\ref{eq:closure:msa}).  For the former $\beta_e=\beta$ so that
$\beta_{ec}= \beta_c$, but one notes that both MSA and SCOZA have
the same configurational internal energy at the critical point just as
in the lattice case.

Now with relation~(\ref{eq:core:renorm}) the internal energy $u_1$ can
no longer be equal to the MSA one at the critical point since the
vertex function $\mu_c$ is different.  According to the relation
\begin{displaymath}
1-\mu_c\beta_{ec}(-\tilde\psi(0)) = 0
\end{displaymath}
a change in $\mu_c$ means a change in $\beta_{ec}$ while $\mu_c K$ in
Eq.~(\ref{eq:scoza:u1}) remains unchanged.  The change in $\mu_c$ thus
has the consequence that the internal energy contribution from $u_1$ at the critical point, $u_{1c}  \propto
1/\beta_{ec} =\mu_c$, is modified, too.

In SCOZA computations the shift in $ T_c$ relative to the mean field
value $ T_c^\mathrm{MF}$ is determined by~(\ref{eq:core:renorm}) to
leading order and not by the MSA $\mu_c$ from~(\ref{eq:core:mu}) since
the former is closer to thermodynamic consistency.  At the same time
the SCOZA energy $u_1$ is dictated by the MSA $\mu_c$.  Thus there
results an inaccuracy in the SCOZA $u_1$ with a corresponding
inaccuracy in the SCOZA $ T_c$.  The reason is that to leading order
the shift down in $ T_c$ relative to $ T_c^\mathrm{MF}$ is
proportional to $u_1$.  In terms of $u_1$, Eq.~(\ref{eq:core:renorm})
is ($\beta_e=\beta$)
\begin{equation} \label{eq:jh:16}
\mu = \mu_c (1+A),\qquad A=-\mu''\beta u_1/\rho.
\end{equation}
However, by solving SCOZA the $u_1$ is changed into the MSA value that
follows from the $\mu_c=\mu_c^\mathrm{MSA}$ of Eq.~(\ref{eq:core:mu})
instead of Eqs.~(\ref{eq:core:renorm}) or~(\ref{eq:jh:16}),
\begin{displaymath}
u_1^\mathrm{MSA} = \frac{\mu_c^\mathrm{MSA}}{\mu_c}\,u_1.
\end{displaymath}
This new value modifies the equation~(\ref{eq:jh:16}) into
\begin{displaymath}
\mu = \mu_c\left(1+A\frac{\mu_c^\mathrm{MSA}}{\mu_c}\right).
\end{displaymath}
When inserted into~(\ref{eq:jh:13}), this gives an additional shift in
$ T_c$ compared with use of Eq.~(\ref{eq:jh:16}), \textit{viz.{}},
\begin{equation} \label{eq:jh:19}
\Delta T_c
= \left(\frac1{1+A} - \frac1{1+A\frac{\mu_c^\mathrm{MSA}}{\mu_c}}\right)
T_c^\mathrm{MF}
\approx - \left(1-\frac{\mu_c^\mathrm{MSA}}{\mu_c}\right) A 
T_c^\mathrm{MF}.
\end{equation}
Now with Eq.~(\ref{eq:core:mu}) we approximate
\begin{displaymath}
\mu \approx \mu_c^\mathrm{MSA} (1+\mu_c K)
\approx \mu_c^\mathrm{MSA} (1+2A),
\end{displaymath}
where the last relation follows from comparison with
Eq.~(\ref{eq:core:renorm}) and using the numerical
value~(\ref{eq:jh:12}) for $\mu''$.  Inserted into
Eq.~(\ref{eq:jh:19}) and expanded this means
\begin{equation} \label{eq:jh:20}
\Delta T_c \approx - A^2  T_c^\mathrm{MF}.
\end{equation}
Evaluations for the Lennard-Jones (LJ) interaction gave
\cite{hoye:dr,hoye::nato}
\begin{displaymath}
A = 1 - \frac{\mu_c}\mu
= 1 - \frac{ T_c}{ T_c^\mathrm{MF}}
= 1 - \frac{0.2545\cdot4}{0.272\cdot4}
= 0.064
\end{displaymath}
so that
\begin{equation} \label{eq:scoza:shift:estimate}
\Delta T_c \approx -4.2\cdot 10^{-3} T_c^\mathrm{MF}
\approx -4.5\cdot 10^{-3} T_c.
\end{equation}

\section{Modified SCOZA direct correlation function}

\label{sec:theory}

The lowering on $ T_c$ given by Eq.~(\ref{eq:jh:20}) can be understood
in terms of a graph expansion.  There is a significant contribution of
short range to $h(r)$ and the direct correlation function $c(r)$ that
is not properly taken care of by previous SCOZA computations.  To
leading order it is connected to the $\mu''$-term of
Eq.~(\ref{eq:core:renorm}).  In terms of graphs, $\mu$ represents the
hypervertex where the graphs are those of the reference system
correlation function plus $\delta(\vec r)$, \textit{i.~e.{}},
$\rho\,\delta(\vec r) + \rho^2\,h_0(r)$.  On two $\rho$-vertices of
these graphs a chain bond $K$ with endpoints or vertices $\mu$, that
can be replaced by $\mu_c$ to leading order, and symmetry factor
$\frac12$ are added.  This contribution is then added to the reference
system correlation function or vertex $\mu$ to obtain the renormalized
vertex $\mu_c$ \cite{hoye:dr,hoye::nato}.

For lattice gases, however, we do not need to consider this term
explicitly as the core condition~(\ref{eq:core:mu}) is then equivalent
to~(\ref{eq:core:renorm}).  To fulfill the former, a constant $c_0$ is
added to the direct correlation function $c(r)$ at $\vec r=0$.  This
can also be regarded as a short range piece of the perturbing
potential such that $\tilde v(k)$ in expression~(\ref{eq:core:mu}) is
replaced by $c_0 + \tilde v(k)$.  The consequence of this is that the
chain bond integral~(\ref{eq:core:mu}) becomes zero and
renormalization~(\ref{eq:core:renorm}) drops out as $c_0$ has already
taken care of it.

For continuum fluids, however, Eq.~(\ref{eq:core:renorm}) can not be
replaced by a $c_0(r)$ only inside hard cores if the core condition is
to be kept; the short range part of it goes outside the core in
contrast to the lattice gas system.  Thus $c(r)$ must have a
contribution of short range outside the hard core (beyond that of the
hard sphere reference system, $ c_{HS}(r)$).  This term will have both
an amplitude and a range.  In a SCOZA scheme this means two unknown
parameters if now the remaining part of $c(r)$ is just its MSA form
outside the core, \textit{i.~e.{}}, the new term replaces the earlier
need for an effective temperature.

Assuming the added short-range piece to the direct correlation to be
of Yukawa form, too, the SCOZA direct correlation function for the
one-Yukawa potential now becomes
\begin{equation} \label{eq:closure}
c(r) = K_1 \frac{e^{-z_1\,(r-1)}}r + K_2 \frac{e^{-z_2\,(r-1)}}r
+  c_{HS}(r),
\qquad r>1.
\end{equation}
The first term can be treated as in MSA, \textit{i.~e.{}}, its
amplitude can be identified with the inverse temperature, $K_1 \equiv
\beta$, while its form coincides with that of the interaction, $z_1
\equiv z$; this also means that $c(r)$ asymptotically coincides with
$-\beta\psi(r)$ for large $r$ as it should except at the critical
point.  $K_2$ and $z_2$, on the other hand, are the two unknown
parameters of the added short range piece to $c(r)$.  Obviously,
conventional SCOZA is recovered when setting $z_2 = z_1 = z$, with
$K_1 + K_2$ playing the role of the effective inverse temperature
$\beta_e$.

The use of an effective parameter $\beta_e$ is common to
conventional SCOZA problems.  So the differences between various SCOZA
problems lie in the actual equations used.  These again depend upon
whether continuum fluids or lattice gases are considered, and further
they depend upon the pair interaction used. However, in the present
case this conventional scheme is modified by which we have two unknown
parameters present in Eq.~(\ref{eq:closure}).   As only one
parameter can be determined in ``standard'' SCOZA, an option is to
keep the range fixed and to make evaluations for various values of
$z_2$ to see how the estimated change in $ T_c$ comes out.
Furthermore, SCOZA allows the contact value of $h(r)$ at the hard core
surface to be obtained so that we can also check consistency with the
virial theorem.  Alternatively, the range parameter $z_2$ can also be
adjusted alongside $K_2$ as a function of $\beta$ and $\rho$ so that
consistency between all three of the energy, virial, and
compressibility routes is obtained throughout the temperature-density
plane.  In this more ambitious SCOZA evaluation first proposed by H\o
ye and Stell \cite{scoza:11} both unknown parameters, $K_2$ and $z_2$,
are supposed to be determined simultaneously.  This scheme thus bears
some resemblance to the version of the Generalized Mean Spherical
Approximation studied in Ref.~\onlinecite{allg:7} that did not,
however, take into account $ c_{HS}(r)$ outside the core.

Yet another possibility is to keep $z_2$ fixed for all densities and
temperatures, choosing it in such a way that consistency with the
virial theorem is achieved at the critical point.  This has the
advantage over full consistency of being much easier to implement
while it can be expected that this simpler version already
incorporates most of the effect of full consistency with the virial
route, at least when using the location of the critical point and the
phase boundary to gauge the relative accuracy and performance of
various liquid state theories as we will do here.

\section{Numerical results}

\label{sec:num}

In our computations we found it advantageous to adopt the Waisman
recipe for the direct correlation function $ c_{HS}(r)$ of the hard
sphere reference system, \textit{i.~e.{}},
\begin{displaymath}
c_{HS}(r) =  K_{HS}\,\frac{e^{- z_{HS}\,(r-1)}}r, \qquad r>1,
\end{displaymath}
with density dependent parameters $ K_{HS}$ and $ z_{HS}$ that
reproduce the Carnahan Starling equation of state
\cite{scoza:6,waisman::1973}.  Just as in conventional SCOZA, the
direct correlation function outside the hard core is then just a sum
of Yukawa terms for which the solution of the OZ equation together
with the core condition, $h(r)=-1$ for $r<1$, is known
semi-analytically by way of a Wiener-Hopf factorization and
Eqs.~(\ref{eq:route:eny}) and~(\ref{eq:route:comp}) can be evaluated
efficiently~\cite{allg:33,allg:28}.

The main change required vis-\`a-vis the conventional theory is the
shift from $K_1$, which acquires an explicit temperature dependence,
to $K_2$ as the adjustable parameter; also, it becomes necessary to
implement the virial route expressions\footnote{\label{fn:arrieta}{}In
this respect it is worth pointing out that the expressions given in
appendix~III of Ref.~\onlinecite{allg:28} are only valid for $K_1 = K_2 =
K_{HS} = \beta$ and thus cannot be used here.  The results of
Ref.~\onlinecite{allg:33}, while not optimized for numerical evaluation, are
free of this restriction.}  that do not normally feature in SCOZA
computations but already appeared in the preliminary evaluation of our
recent adaptation of SCOZA to molecules with soft cores \cite{ar:21}.

For an interaction composed of a hard core of unit diameter with a
perturbing tail $\psi(r)$, the virial theorem,
\begin{displaymath}
  \frac{\beta P}\rho = 1 + \frac{2 \pi}3 \rho \*
   \left[ 1 + h(1+)
    - \beta \int_1^\infty r^3\,(h(r)+1)\,\psi'(r)\,\mathrm{d}r\right],
\end{displaymath}
relates the pressure $P$ both to the contact value $h(1+)$ of the pair
correlation function and to an averaged inter-particle force away from
contact.  In the case of a direct correlation function of multi-Yukawa
form outside the core, the same steps that lead to the energy route
$u$ and the compressibility route $a$, Eqs.~(\ref{eq:route:eny})
and~(\ref{eq:route:comp}), also provide us with explicit expressions
for the Laplace transform $G(s)$ of $r+r\,h(r)$ \cite{allg:33} from
which the contact value can be worked out \cite{allg:28}.  If, on the
other hand, the interaction is a sum of Yukawa terms with inverse
length scale parameters $z_i$, it is easily seen that the integral on
the right hand side of the relation above reduces to a sum of linear
combinations of $G(s)$ and $\mathrm{d}G(s)/\mathrm{d}s$ with $s$ set
to each of the $z_i$.  As both conditions are met in our case, the
virial pressure can be evaluated numerically no less efficiently than
the other thermodynamic quantities commonly entering SCOZA.

Implementation of the theory outlined in section \ref{sec:theory}
started from our re-imple\-men\-tation of conventional SCOZA, which
has been developed with a view to great conceptual simplicity,
modularity, and flexibility and has turned out to be a convenient
testbed for modifications of both the theory and its numerical
realization.  Methodically it largely follows the work of
Sch\"oll-Paschinger \cite{paschinger:dr}; the main differences are the
avoidance of general non-linear solving routines through the
systematic use of linearizations (which also eliminates the need for
cumbersome implicit differentiations and incidentally brings about a
significant speed advantage) and an improved implementation of the
artificial spinodal boundary condition (\textit{v.~i.{}}) that is more
efficient computationally and does not restrict the estimated spinodal
to the discretization grid, nor does it use the ill-founded criterion
of minimal width of the spinodal.

For the computations below, we solved the SCOZA
PDE~(\ref{eq:scoza:pde}) by an iterated predictor-corrector method on
density grids of $N_\rho = 10^3$ density steps spanning the range $0
\le \rho \le \rho_{\rm max} = 1$ with the ideal gas ($u=0$) and the
MSA ($K_2=0$) as boundary conditions at $\rho=0$ and $\rho= \rho_{\rm
max}$, respectively.  The inverse temperature $\beta$ was turned on
from $\beta=0$, where the hard sphere reference fluid provided the
initial conditions, down to the critical point with temperature steps
decreasing upon approaching the critical point in approximate
proportionality to $\min_\rho \sqrt{a(\beta,\rho)}$.  Below the
critical point the temperature steps increased again in such a way
that the shift of the spinodal per integration step was limited to
less than one density step.  Below the critical point there is a
region where the analytical solution of the OZ equation becomes
invalid \cite{pastore::1988}, and even a fully numerical procedure
\cite{scoza:26} that does not rely on the analytical result eventually
runs into problems of numerical convergence \cite{esp::pc}.  Either
way, it proves necessary to remove part of the solution from the PDE's
domain through the imposition of an artificial boundary condition at
the spinodal.  In our implementation we do so by estimating its
location at the new temperature from the solution of the PDE at the
previous temperature on nearby grid points and forcing the internal
energy there to be compatible with $a=0$.  Pressure and chemical
potential are obtained by thermodynamic integration at fixed density,
starting from the hard core reference fluid at $\beta=0$.  This allows
the binodal to be found by locating densities $ \rho_v(\beta)$ and $
\rho_l(\beta)$ of the coexisting vapour and liquid phases of equal
pressure, $P( \rho_v,\beta) = P( \rho_l,\beta)$, and chemical
potential, $\mu( \rho_v,\beta) = \mu(\rho_l,\beta)$.  The critical
point is identified with the locus where the liquid and vapour
branches of the spinodal and the binodal meet.

This computation is then repeated for different values of $z_2$.  To
illustrate the dependency of the critical parameters on $z_2$, let us
consider the potential with $z=1.8$ that is known to exhibit
thermodynamics roughly equivalent to the LJ interaction.  As can be
seen from Fig.~\ref{fig:z2:betaC}, the critical temperature obtained
with $z_2> z_1 = z$ is indeed a bit lower than that for conventional
SCOZA ($z_2 = z_1 = z$), and typical shifts are comparable to the
estimate~(\ref{eq:scoza:shift:estimate}) obtained using results for
the LJ interaction.

As we solve the SCOZA PDE, we compare the pressure obtained from
thermodynamic integration to the virial route result.  In general, for
every value of $z_2$ there will be lines along which consistency
between SCOZA and the virial theorem occurs; this is illustrated,
again for the case of $z=1.8$, in Fig.~\ref{fig:phase}.  As can be
seen, the geometry of these lines is rather complex, especially below
the critical point.  We expect this distribution of $z_2$ values to
resemble quite closely the function $z_2(\beta,\rho)$ that would be
obtained with the more general version of the theory imposing
consistency with the virial route throughout the PDE's domain,
\textit{cf.{}} section~\ref{sec:theory}.

In the present contribution, however, we consider a simpler version of
SCOZA where $z_2$ is determined from consistency at the critical point
only.  Numerically the search for this optimal value of $z_2$ calls
for repeated solutions of the SCOZA PDE and might thus seem to be
exceedingly burdensome.  In contrast to, \textit{e.~g.{}},
Ref.~\onlinecite{ar:20} where this is indeed a concern, however, we can use
the semi-analytic solution of the OZ equation.  In combination with a
very small number of Yukawa terms in Eq.~(\ref{eq:closure}) the
integration of the PDE down to the critical temperature is
sufficiently fast so that the search for $z_2$ does not constitute a
problem.

In table \ref{tab:comparison} we list the results both of the version
of SCOZA outlined in section~\ref{sec:theory} ($z_2 > z_1 = z$) and
for conventional SCOZA ($z_2 = z_1 = z$) for various potentials, and
we compare them to moderately recent simulations: Monte Carlo (MC)
results obtained with $N=108$ particles for $z \in \{3.9, 7, 25,
100\}$ \cite{hc-y:8}; Gibbs Ensemble MC (GEMC) data for $z \in \{1.8,
3, 4\}$ ($N=108$, \cite{hc-y:6}) and for $z \in \{3.9, 7\}$ ($N=216$,
\cite{hc-y:7}); as well as the results of Grand Canonical MC (GCMC)
simulations in combination with Finite Size Scaling (FSS) analysis for
$z = 1.8$ \cite{scoza:6}.  The same data are also displayed in
Fig.~\ref{fig:comparison}.  It should be noted that the liquid-gas
transition considered here is metastable with respect to the solid for
rather short-ranged potentials, \textit{viz.{}}, for $z \gtrsim 6$
\cite{hc-y:7}; in particular, the highest $z$ values for which
simulations are available, \textit{viz.{}}, $z\in\{25, 100\}$
correspond to interactions for which the fluid-fluid transition is
buried deep inside the metastable region \cite{hc-y:8}.

First limiting ourselves to moderately short-ranged potentials ($z \le
10$), a comparison of conventional SCOZA ($z_2 = z_1 = z$) and of our
modified SCOZA proposed here ($z_2 \ne z_1 = z$) shows two general
trends: As expected, the critical temperature is shifted to lower
values; for the LJ value of $z=1.8$ this difference is about $0.7\,\%$
and thus well compatible with the
estimate~(\ref{eq:scoza:shift:estimate}).  Furthermore, the shift in
critical temperature is also accompanied by a small decrease of the
critical density.  Both effects become more pronounced as $z$ becomes
larger.  The available simulations confirm that the shift in $ T_c$ is
certainly in the right direction, and the agreement is excellent
considering the uncertainties of the simulation studies that generally
cite non-overlapping confidence intervals for the same interaction.
For some values of $z$ the prediction of our modified SCOZA falls
right into the confidence interval of some study, as is the case for
$z=1.8$ and the sophisticated analysis of Ref.~\onlinecite{scoza:6} or for
$z=4$ and the GEMC result of Ref.~\onlinecite{hc-y:6}.  For the nearby value
of $z=3.9$ deemed relevant to the fullerene C${}_{60}$, on the other
hand, two different studies \cite{hc-y:8,hc-y:7} agree that our $ T_c$
is too high, if by significantly different amounts, whereas our
prediction is right between the results of the same two references for
$z=7$.  For $z=3$, our modified SCOZA is marginally compatible with
the GEMC\ results.  By way of contrast, conventional SCOZA
($z_2=z_1=z$) yields too high a critical temperature in all cases.
This situation is summarized graphically in Fig.~\ref{fig:comparison};
to judge by $ T_c$, the variant of SCOZA studied in the present
contribution thus constitutes a genuine progress over its conventional
form, and the available simulations are not sufficiently accurate to
point out systematic deficiencies in the critical temperatures
predicted.  As for the critical density, both old and new SCOZA are
well compatible with the results for $z\in\{3, 3.9, 4\}$.  Only for
$z=1.8$ are we able to differentiate between the two versions of the
theory: the modified SCOZA is in perfect agreement with the the
extremely narrow confidence interval obtained by way of GCMC and FSS
methods whereas conventional SCOZA lies somewhat outside this range,
thus hinting at the superiority of our modifications for the critical
density, too.

As the potential becomes more and more short-ranged, however,
discrepancies start to appear: For one, our modifications always bring
about a shift of the critical density to lower values, but at $z=7$
the GEMC $\rho_c$ is already higher than that of conventional SCOZA.
Similarly, the critical temperature is always lowered, but the MC\
simulations for the two highest values of $z$ listed in
Tab.~\ref{tab:comparison}, \textit{viz.{}}, $z\in\{25, 100\}$ yield
critical temperatures that are far greater even than those of
conventional SCOZA; furthermore, consistency with the virial theorem
does not seem achievable at all for $z=100$, presumably due to
numerical difficulties with the evaluation of the virial pressure
(\textit{cf.{}}\ footnote~\ref{fn:arrieta}).  Clearly, more work is
needed to clarify the situation at these extremely high values of $z$
that are, however, far outside the range of stability of the
liquid-vapour transition and thus of limited interest for a liquid
state theory.

\section{Summary}

\label{sec:summary}

In the present contribution we have considered in some depth the
question of thermodynamic consistency between MSA and SCOZA (that
builds upon MSA but incorporates consistency between the energy and
compressibility routes) on the one hand, and the virial route on the
other hand.  The basis for our investigations is formed by an analysis
of the solution of the MSA closure in the presence of the core
condition in the light of renormalized $\gamma$-ordering that
highlights a crucial difference between the lattice and the continuum
cases.  For the latter it is seen that a short-range piece must be
added to the direct correlation function, and it is this term that we
propose to be adjusted instead of the usual potential term.  This
scheme has been implemented numerically in a simplified version where
only the amplitude of this contribution to $c(r)$ is varied throughout
the $(\beta,\rho)$-plane according to the usual SCOZA procedure
implementing consistency between the energy and compressibility routes
only whereas its range is fixed by imposing consistency with the
virial route at the critical point only.  Comparison with the results
of various simulation studies shows that these modifications bring
about significant improvements over conventional SCOZA; particularly
encouraging is the perfect agreement with a sophisticated and
extremely accurate study combining GCMC simulations with FSS analysis
for $z=1.8$ \cite{scoza:6}.

If all our evaluations have been restricted to the hard-core Yukawa
class of interactions, the question arises naturally whether this
approach can be applied to other types of potentials, too.  In this
respect it is clear that nothing prevents us from adding some Yukawa
term to the MSA direct correlation function for, say, a Lennard-Jones
potential and using the same strategies for evaluating its range
parameter and solving the modified SCOZA PDE.  At the same time,
however, the form of the Yukawa potential is special in that it is
most attractive right at the core whereas more realistic interactions
reach the maximum of the depth of the potential only at greater
distances.  For these a contribution of Yukawa form to the direct
correlation function hardly seems appropriate.  So clearly more work
is needed before the results presented here can be transferred to
other cases.  Nevertheless, we expect the strategy of using the virial
theorem (that does not traditionally feature in SCOZA at all) as a
means of gauging the fluid structure close to the repulsive core to
remain the key for fixing the form of the adjustable part of the
direct correlation function.  This strategy, it should be noted, also
featured prominently in our earlier work on SCOZA for molecules with
soft cores \cite{ar:21}.

\section*{Acknowledgments}

AR gratefully acknowledges financial support for part of this work
from \textit{Fonds zur F\"orderung der wissenschaftlichen Forschung
(FWF)} under project~J2380-N08.

\section*{TABLE CAPTION}

TABLE 1: {Comparison of the critical temperatures and densities for
hard-core Yukawa fluids of various inverse range parameters $z$.
SCOZA results were obtained as described in section~\ref{sec:num},
with $z_2$ determined from consistency with the virial route as
proposed in section~\ref{sec:theory}, or set equal to $z$ as in
conventional SCOZA.  The simulation results are taken from the
literature.}

\section*{FIGURE CAPTIONS}

FIGURE 1: Variation of the critical temperature $ T_c$ with the
inverse range parameter $z_2$ of the adjustable part of the direct
correlation function for the potential with $z=1.8$.  The $z_2$ is
varied from $z_2 = z$, corresponding to conventional SCOZA ($ T_c =
T_c^S$), up to $z_2 = 600$.

FIGURE 2: Loci of consistency with the virial route for the hard-core
Yukawa potential with $z=1.8$ and various values of $z_2$.
Consistency is achieved at the critical point for $z_2 = z_2^* =
7.57145$.  The binodal obtained in this case is shown by the dotted
line.  This figure combines the results of several runs of our SCOZA
implementation during each of which $z_2$ was kept fixed as opposed to
a more general version of the theory where $z_2$ would be adjusted
during a single run.

FIGURE 3: Critical temperatures $ T_c$ for hard-core Yukawa potentials
with varying inverse length scale $z$ as computed by conventional
SCOZA (dashed line) and by various simulations relative to the
prediction $ T_c^*$ of our modified version of SCOZA.  For the
simulations, the crosses mark the value of $ T_c$, the vertical bars
indicate the uncertainties where available, and the labels identify
the source: ``A'' --- GCMC, FSS (Ref.~\onlinecite{scoza:6}); ``B'' ---
GEMC ($N=108$, Ref.~\onlinecite{hc-y:6}); ``C'' --- MC ($N=108$,
Ref.~\onlinecite{hc-y:8}); ``D'' --- GEMC ($N=216$,
Ref.~\onlinecite{hc-y:7}); cf.~table~\ref{tab:comparison}.

\begin{table}
{\singlespacing
\begin{tabular}{|l|l|l|l|}
\hline
$z$ & $ T_c$ & $\rho_c$ & method \\
\hline
1 & 2.51814 & 0.279 & SCOZA, \(z_2 = z\) \\
\cline{2-4}
& 2.51388 & 0.279 & SCOZA, \(z_2 = 5\).95560 \\
\hline
1.8 & 1.21869 & 0.3145 & SCOZA, \(z_2 = z\) \\
\cline{2-4}
& 1.212(2) & 0.312(2) & GCMC, FSS (Ref.~\onlinecite{scoza:6}) \\
\cline{2-4}
& 1.21013 & 0.312 & SCOZA, \(z_2 = 7\).57145 \\
\cline{2-4}
& 1.177(5) & 0.313(13) & GEMC ($N=108$, Ref.~\onlinecite{hc-y:6}) \\
\hline
2 & 1.08811 & 0.323 & SCOZA, \(z_2 = z\) \\
\cline{2-4}
& 1.07877 & 0.320 & SCOZA, \(z_2 = 8\).03331 \\
\hline
3 & 0.74025 & 0.359 & SCOZA, \(z_2 = z\) \\
\cline{2-4}
& 0.72860 & 0.352 & SCOZA, \(z_2 = 10\).56167 \\
\cline{2-4}
& 0.715(11) & 0.375(27) & GEMC ($N=108$, Ref.~\onlinecite{hc-y:6}) \\
\hline
3.9 & 0.60200 & 0.387 & SCOZA, \(z_2 = z\) \\
\cline{2-4}
& 0.58980 & 0.373 & SCOZA, \(z_2 = 13\).08509 \\
\cline{2-4}
& 0.5714 & --- & MC ($N=108$, Ref.~\onlinecite{hc-y:8}) \\
\cline{2-4}
& 0.549(3) & 0.37(2) & GEMC ($N=216$, Ref.~\onlinecite{hc-y:7}) \\
\hline
4 & 0.59108 & 0.390 & SCOZA, \(z_2 = z\) \\
\cline{2-4}
& 0.57889 & 0.375 & SCOZA, \(z_2 = 13\).37749 \\
\cline{2-4}
& 0.576(6) & 0.377(21) & GEMC ($N=108$, Ref.~\onlinecite{hc-y:6}) \\
\hline
5 & 0.50885 & 0.415 & SCOZA, \(z_2 = z\) \\
\cline{2-4}
& 0.49678 & 0.392 & SCOZA, \(z_2 = 16\).43066 \\
\hline
6 & 0.45624 & 0.437 & SCOZA, \(z_2 = z\) \\
\cline{2-4}
& 0.44390 & 0.402 & SCOZA, \(z_2 = 19\).75802 \\
\hline
7 & 0.41881 & 0.454 & SCOZA, \(z_2 = z\) \\
\cline{2-4}
& 0.411(2) & 0.50(2) & GEMC ($N=216$, Ref.~\onlinecite{hc-y:7}) \\
\cline{2-4}
& 0.40530 & 0.403 & SCOZA, \(z_2 = 23\).44057 \\
\cline{2-4}
& 0.4000 & --- & MC ($N=108$, Ref.~\onlinecite{hc-y:8}) \\
\hline
8 & 0.38977 & 0.463 & SCOZA, \(z_2 = z\) \\
\cline{2-4}
& 0.37446 & 0.397 & SCOZA, \(z_2 = 27\).54452 \\
\hline
9 & 0.36574 & 0.465 & SCOZA, \(z_2 = z\) \\
\cline{2-4}
& 0.34842 & 0.386 & SCOZA, \(z_2 = 32\).09389 \\
\hline
10 & 0.34497 & 0.462 & SCOZA, \(z_2 = z\) \\
\cline{2-4}
& 0.32578 & 0.372 & SCOZA, \(z_2 = 37\).08103 \\
\hline
25 & 0.2353 & --- & MC ($N=108$, Ref.~\onlinecite{hc-y:8}) \\
\cline{2-4}
& 0.18706 & 0.307 & SCOZA, \(z_2 = z\) \\
\cline{2-4}
& 0.15829 & 0.170 & SCOZA, \(z_2 = 165\).80499 \\
\hline
100 & 0.1538 & --- & MC ($N=108$, Ref.~\onlinecite{hc-y:8}) \\
\cline{2-4}
& 0.06059 & 0.07 & SCOZA, \(z_2 = z\) \\
\hline
\end{tabular}
}
\caption{}
\label{tab:comparison}
\end{table}

\begin{figure}[p]\vbox{\noindent\includegraphics{fig1.mps}
\caption{}
\label{fig:z2:betaC}}
\end{figure}

\begin{figure}[p]\vbox{\noindent\includegraphics{fig2.mps}
\caption{}
\label{fig:phase}}
\end{figure}

\begin{figure}[p]\vbox{\noindent\includegraphics{fig3.mps}
\caption{}
\label{fig:comparison}}
\end{figure}

\begin{thebibliography}{99}

\bibitem{scoza:4}D.\ Pini, G.\ Stell, J.~S.\ H\o ye, Int.\
J.\ Thermophys.\ {\bf 19}, 1029 (1998).
\bibitem{scoza:6}D.\ Pini, G.\ Stell, N.~B.\
Wilding, Mol.\ Phys.\ {\bf 95}, 483 (1998).
\bibitem{scoza:20}A.\ Borge, J.\ S.\ H\o ye, J.\ Chem.\ Phys.\ {\bf 108},
4516 (1998).
\bibitem{scoza:8}D.\ Pini, G.\ Stell, R.\ Dickman, Phys.\ Rev.\ E\ {\bf
57}, 2862 (1998).
\bibitem{scoza:21}J. S. H\o ye, A. Borge, J. Chem.\ Phys.\ {\bf 108}, 8830
(1998).
\bibitem{scoza:26}E.\ Sch\"oll-Paschinger, A.\ L.\
Benavides, R.\ Casta\~neda-Priego, J.\ Chem.\ Phys.\ {\bf 123}, 234513
(2005).
\bibitem{scoza:25}J.\ S.\ H\o ye, D.\ Pini, G.\
Stell, Physica A\ {\bf 279}, 213 (2000).
\bibitem{scoza:1}R.\ Dickman, G.\ Stell, Phys.\ Rev.\ Lett.\ {\bf 77}, 996
(1996).
\bibitem{hc-y:6}E.\ Lomba, N.\ G.\ Almarza, J.\ Chem.\
Phys.\ {\bf 100}, 8367 (1994).
\bibitem{hoye:dr}J.~S.\ H\o ye, {\it A study of phase
transitions in systems with long-range forces}, PhD\ thesis, NTH Trondheim
(1973).
\bibitem{hoye::nato}J.~S.\ H\o ye, {\it Equation of state and
critical behavior of simple fluid models}, in: C.\
Caccamo et al.\ (eds.), Proceedings of the NATO Advanced Study
Institute on {\it New Approaches to old and new problems in liquid
state theory}, Patti Marina, Messina, Italy, July 7--17, 1998,
Dordrecht (Kluwer) 1999, 9--29.
\bibitem{hemmer:1964}P.~C.\ Hemmer, J.\ Math.\
Phys.\ {\bf 5}, 75 (1964).
\bibitem{lsb:1965}J.~L.\ Lebowitz, G.\ Stell, S.\
Baer, J.\ Math.\
Phys.\ {\bf 6}, 1282 (1965).
\bibitem{scoza:11}J.~S.\ H\o ye, G.\ Stell, Mol.\ Phys.\ {\bf 52}, 1071
(1984).
\bibitem{allg:7}C.\ Caccamo, G.\ Pellicane, D.\ Costa,
D.\ Pini, G.\ Stell, Phys.\ Rev.\ E\ {\bf 60}, 5533 (1999).
\bibitem{waisman::1973}E.\ Waisman, Mol.\ Phys.\ {\bf 32}, 1627 (1973).
\bibitem{allg:33}J.\ S.\ H\o ye, L.\ Blum, J.\ Stat.\
Phys.\ {\bf 16}, 399 (1977).
\bibitem{allg:28}E.\ Arrieta, C.\ Jedrzejek, K.\ N.\
Marsh, J.\ Chem.\ Phys.\ {\bf 95}, 6806 (1991).
\bibitem{ar:21}J.\ S.\ H\o ye,
A.~Reiner, J.\ Chem.\ Phys.\ {\bf 125}, 104503 (2006).
\bibitem{paschinger:dr}E.\ Sch\"oll-Paschinger, {\it Phase
behavior of simple fluids and their mixtures}, PhD\ thesis, Technische
Universit\"at Wien (2002).
\bibitem{pastore::1988}G.\ Pastore, Mol.\ Phys.\ {\bf 55}, 187 (1988).
\bibitem{esp::pc}E.\ Sch\"oll-Paschinger, private
communication.
\bibitem{ar:20}E.\
Sch\"oll-Paschinger, A.~Reiner, J.\ Chem.\ Phys.\ {\bf 125}, 164503 (2006).
\bibitem{hc-y:8}M.\ Dijkstra, Phys.\ Rev.\ E\ {\bf 66}, 021402 (2002).
\bibitem{hc-y:7}M.\ H.\ J.\ Hagen, D.\ Frenkel, J.\ Chem.\ Phys.\ {\bf
101}, 4093 (1994).

\end{thebibliography}
\end{document}